\documentclass[
reprint,
superscriptaddress,
amsmath,amssymb,
aps,
prx,
longbibliography]{revtex4-1}

\usepackage{graphicx}% include figure files
\usepackage{dcolumn}% align table columns on decimal point
\usepackage{bm}% bold math
\usepackage[colorlinks=true,urlcolor=blue,linkcolor=blue,citecolor=blue]{hyperref}% add hypertext capabilities
\usepackage{ulem}

\newcommand{\customref}[2]{\hyperref[#1]{\ref*{#1}#2}}

\usepackage{xcolor}

\AtBeginDocument{%
    \newwrite\bibnotes
    \def\bibnotesext{Notes.bib}
    \immediate\openout\bibnotes=\jobname\bibnotesext
    \immediate\write\bibnotes{@CONTROL{REVTEX41Control}}
    \immediate\write\bibnotes{@CONTROL{%
    apsrev41Control,author="08",editor="1",pages="1",title="0",year="1"}}
     \if@filesw
     \immediate\write\@auxout{\string\citation{apsrev41Control}}%
    \fi
}%

\begin{document}

\title{Many-body localization near the critical point}

\author{Alan Morningstar}
\affiliation{Department of Physics, Princeton University, Princeton, NJ 08544, USA}

\author{David A. Huse}
\affiliation{Department of Physics, Princeton University, Princeton, NJ 08544, USA}
\affiliation{Institute for Advanced Study, Princeton, NJ 08540, USA}

\author{John Z. Imbrie}
\affiliation{Department of Mathematics, University of Virginia, Charlottesville, VA 22904, USA}
%\email{imbrie@virginia.edu}

\date{\today}

\begin{abstract}

We examine the many-body localization (MBL) phase transition in one-dimensional quantum systems with quenched randomness and short-range interactions.  Following recent works, we use a strong-randomness renormalization group (RG) approach where the phase transition is due to the so-called avalanche instability of the MBL phase.  We show that the critical behavior can be determined analytically within this RG.  On a rough {\it qualitative} level the RG flow near the critical fixed point is similar to the Kosterlitz-Thouless (KT) flow as previously shown, but there are important differences in the critical behavior.  Thus we show that this MBL transition is in a new universality class that is different from KT.  The divergence of the correlation length corresponds to critical exponent $\nu\rightarrow\infty$, but the divergence is weaker than for the KT transition.

\end{abstract}

\maketitle

% \tableofcontents

\section{Introduction\label{sec:introduction}}

The many-body localization (MBL) phase transition is a dynamical quantum phase transition in the entanglement and thermalization properties of highly-excited quantum many-body systems~\cite{Nandkishore-Huse2014a,Abanin-Serbyn2018}. The isolated system's unitary dynamics may be given by a time-independent Hamiltonian, or by a Floquet operator due to a Hamiltonian that is periodic in time.  This phase transition is between the thermal phase where the eigenstates of the system's dynamics obey the Eigenstate Thermalization Hypothesis (ETH)~\cite{Deutsch1991,Srednicki1994,Rigol-Olshanii2009,Dalessio-Rigol2016} and thus are volume-law entangled, and the MBL phase where the eigenstates do not obey the ETH and are only area-law entangled.  In the thermal phase the system is able to function as a thermal reservoir and bring its subsystems to thermal equilibrium at late times via its own unitary dynamics, while in the MBL phase the system instead remains localized in a state ``near'' the initial state. This can be explained by the existence of an extensive set of local integrals of motion (LIOMs) in the MBL phase~\cite{Serbyn-Abanin2013a,Serbyn-Abanin2013b,Huse-Oganesyan2014,Ros-Scardicchio2015,Chandran-Vidal2015}, which cannot be formed in the thermal phase. The details of how these LIOMs delocalize at the transition are still not well understood, and this is part of why MBL transitions remain an unsettled topic.

Broadly speaking, investigations of the nature of the MBL phase transition have been either numerical or renormalization group studies.  The numerical studies are mostly limited to either rather small one-dimensional systems or to short time scales.  For the models with quenched randomness, when the eigenstate data are fit to conventional finite-size scaling for a continuous phase transition~\cite{Kjall-Pollmann2014,Luitz-Alet2015,Khemani-Huse2017b}, the resulting correlation length exponent $\nu$ strongly violates the Chayes, {\it et al.} inequality ($\nu \ge 2$)~\cite{Chandran-Oganesyan2015,Chayes-Spencer1986}.  This indicates that the system sizes being studied are too small to see the correct asymptotic critical behavior, and trends with size consistent with this have been seen~\cite{Khemani-Huse2017a}.  On the other hand, numerical finite-size scaling studies of spin systems where the many-body localization is instead due to a nonrandom quasiperiodic field (and thus not subject to the Chayes, {\it et al.} inequality) may be producing fairly good estimates of the critical behavior for those systems~\cite{Khemani-Huse2017a,Agrawal-Vasseur2020}.

\begin{figure}
\centering
\includegraphics{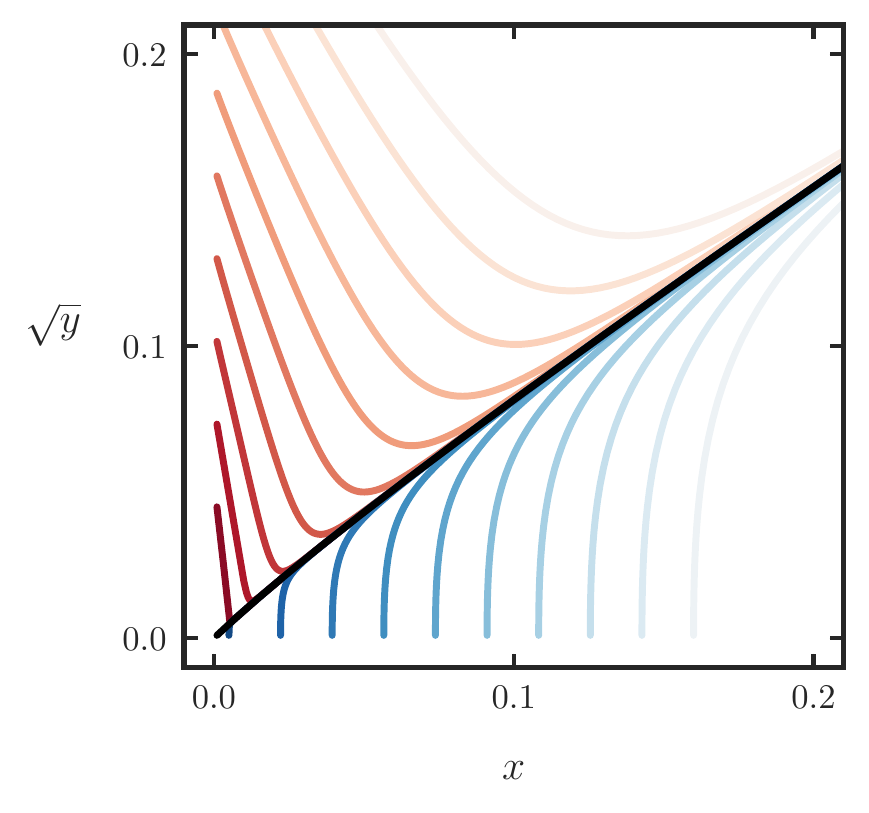}
\caption{\textbf{The RG flow near the critical point.} Flow lines for the system defined by Eqs. (\ref{2.4}) and (\ref{3.2}), with $\sqrt{y}$ plotted on the vertical axis because the separatrix is asymptotic to $y=x^2$ as $x\to 0$.\label{fig:2d_flow}}
\end{figure} 

There have been a number of renormalization group (RG) studies of the phase transition in one-dimensional systems with quenched randomness.  Some, like the present study, assume that the coarse-grained system can be approximated as consisting of distinct locally thermalizing regions and locally insulating regions~\cite{Vosk-Altman2015,Zhang-Huse2016,Goremykina-Serbyn2018,Morningstar-Huse2019}, while others start more microscopically~\cite{Potter-Parameswaran2015,Dumitrescu-Potter2017,Thiery-DeRoeck2017a,Thiery-DeRoeck2017b,Dumitrescu-Vasseur2018}.  Effectively, these are all {\it functional} RGs, with the probability distributions of the local properties of the system being renormalized.  In the earlier works the resulting functional RGs were implemented numerically and essentially fit to standard one-parameter power-law scaling forms~\cite{Vosk-Altman2015,Potter-Parameswaran2015,Dumitrescu-Potter2017}.  

More recently it was argued by Goremykina \textit{et al.}~\cite{Goremykina-Serbyn2018} and Dumitrescu \textit{et al.}~\cite{Dumitrescu-Vasseur2018}
%More recently it was argued by Goremykina, Vasseur and Serbyn~\cite{Goremykina-Serbyn2018} 
that the RG flow is actually Kosterlitz-Thouless-like, with the MBL phase being governed by a fixed line, and the phase transition being governed by an unstable fixed point at the end of that fixed line.  They also showed that the numerical results from the earlier RGs are quite consistent with this scenario, even though they had been instead fit to one-parameter scaling. The conventional scaling assumptions used for exact calculations on small systems are also starting to be revisited~\cite{Suntajs-Vidmar2020,Laflorencie-Mace2020}. In Ref.~\cite{Goremykina-Serbyn2018},
%that paper
certain approximations were made to make the RG analytically tractable, and within those approximations they found that the MBL transition is in the Kosterlitz-Thouless (KT) universality class.
%~\cite{Goremykina-Serbyn2018}.  
Subsequent studies put in more of the correct physics and implemented the RGs numerically, both finding RG flows that are indeed {\it qualitatively} KT-like, but these numerical studies were not able to look in enough detail at the flows to determine whether or not they differ in universality class from KT~\cite{Dumitrescu-Vasseur2018,Morningstar-Huse2019}. What we show in the present paper is that the RG flow near the unstable critical fixed point can be understood analytically and it produces a phase transition that differs in universality class from Kosterlitz-Thouless, even though the RG flows are qualitatively similar to the KT flows. We thus propose a new universality class for the MBL transition in one-dimensional systems with quenched randomness.

Since we work within a certain coarse-grained description, this universality class applies to systems for which that description is appropriate.
%that do coarse-grain to a description that agrees with what we assume: 
Near the critical point, after coarse-graining to 
%at a 
large enough %cutoff 
lengthscales, we assume that the system consists of regions that are locally insulating (I-blocks) or thermalizing (T-blocks), with most of the system being insulating and the T-blocks being of random lengths and randomly located 
%along the chain
in space.  We do know of one class of MBL phase transitions for which this is not true, namely those where the bare model has no quenched randomness, such as models with quasiperiodicity~\cite{Agrawal-Vasseur2020}. It seems that there are two possible scenarios where a non-fine-tuned system with short-range-correlated quenched randomness might have a MBL transition in a different universality class from the one we describe: (1) The MBL phase might have some other instability that causes a transition before the avalanche instability that we are treating. We are not aware of any concrete proposal along these lines, although such a possibility has very recently been suggested in Ref.~\cite{Suntajs-Vidmar2019}; (2) The description of the avalanche instability that we assume, following previous works~\cite{DeRoeck-Huveneers2017,Morningstar-Huse2019}, might need to be revised in some of its details, and those changes would have to be relevant to the universal features of the RG flow in order to change the universality class.  

The remainder of this paper is organized as follows. In Sec. \ref{sec:recursion}, we begin by defining two key RG parameters, which are variations on the ones used in the abovementioned works. One of these ($x$) relates to the decay rate for interactions across insulating regions, and we work with an approximation of spatially uniform $x$. The other ($y$) relates to the density of thermal regions along the line.  Then we show that a basic assumption about the least-unlikely way to generate thermal regions of a given length  leads to a simple recursion relation---see Eq.~(\ref{2.3}) below---that puts strong constraints on how these parameters evolve under the RG. Consequently, any reasonable functional RG satisfying these assumptions  can be  approximated by a specific two-parameter flow---see Eqs. (\ref{2.4}) and (\ref{3.2}) below. In Sec. \ref{sec:critical}, we obtain some analytical results for this proposed RG flow.
%these equations. 
First, the length scale for departure from criticality diverges as 
$\delta_0^{- \log_2 \log_2 \delta_0^{-1}}$, 
where $\delta_0$ is the initial 
%departure 
displacement from the separatrix. Note that the form of this divergence is quite different from the one seen for the standard KT flow. Evidently, $\nu = \infty$, and 
%as this is well above
thus the 
%lower bound of 2 from the
Chayes, \textit{ et al.} inequality is satisfied, so we are able to justify the neglect of spatial variations in $x$.
We also obtain an expression for the fractal exponent of locally thermal regions in the MBL phase near the transition (in agreement with Ref.~\cite{Morningstar-Huse2019}). In Sec. \ref{sec:concrete_RG}, we show that the recursion relation first introduced in Sec. \ref{sec:recursion} through more general arguments can be derived via controlled approximations from a particular functional RG, specifically the one from Ref.~\cite{Morningstar-Huse2019}, by enforcing a spatially uniform $x$. In Sec. \ref{sec:more_accurate}, we show how to obtain a more accurate flow equation by introducing a third parameter and finding the stable manifold in that expanded space. In so doing, we confirm the validity of approximations made earlier in deriving the two-parameter flow. We finish by providing a summary of our results and some concluding remarks.

\section{A recursion for T-block rates\label{sec:recursion}}

When considering the MBL phase transition in one dimension, a real-space RG description has emerged, in which the line is populated by a collection of T-blocks (thermalized blocks) of length $ \ge \Lambda$, where $\Lambda$ is a cutoff length scale.  The eigenstates are assumed to be well thermalized and volume-law entangled within these T-blocks, while the regions in between the T-blocks are insulating for this cutoff.  As the cutoff is increased ($\Lambda \rightarrow \Lambda + d\Lambda$), pairs of adjacent T-blocks satisfying certain %($\Lambda$-dependent) 
proximity conditions are able to thermalize the insulating region between them, so they are combined into a longer T-block.  The remaining T-blocks with lengths in $[\Lambda, \Lambda+d\Lambda]$ are sufficiently isolated so as to be ``erased", \textit{i.e}., absorbed into the surrounding insulating regions.  As explained in Ref.~\cite{Imbrie2016}, if the erasure of T-blocks predominates over the process of T-block combination in the limit of large $\Lambda$, then the system is in the MBL phase or at the phase transition.  If combination predominates, then the system is in the thermal phase. 

A key insight of recent years has been the avalanche mechanism~\cite{Thiery-DeRoeck2017b,DeRoeck-Huveneers2017,Luitz-DeRoeck2017,DeRoeck-Imbrie2017,Goihl-Krumnow2019,Crowley-Chandran2019,Gopalakrishnan-Huse2019}. The insulating regions are characterized by a decay length $\zeta$ for the exponential decay of interactions with distance.  When a T-block is erased, thus combining two adjacent insulating regions, the decay of interactions is interrupted across its former extent, which leads to a reduction in the characteristic decay rate $\zeta^{-1}$ for the new larger insulating region.  When the decay rate reaches a critical value, $\zeta_\text{c}^{-1}$, a finite T-block can thermalize an arbitrarily large insulating region.  Let us choose our unit of length so that $\zeta_\text{c} =1$, and then we may let $x = \zeta^{-1} - 1$ denote the excess decay over this critical rate. The parameter $x$, which characterizes the insulating regions, is then one of two key RG parameters governing the transition~\cite{Morningstar-Huse2019,Dumitrescu-Vasseur2018}.  This parameter $x$ will get renormalized, so more precisely we could denote it by $x_{\Lambda}$ to make this cutoff dependence explicit.  But for brevity we will generally leave off the subscript $\Lambda$.  If the system is within the MBL phase, $x$ remains strictly positive, flowing to a nonzero limiting value as $\Lambda\rightarrow\infty$, while at the phase transition and in the thermal phase $x$ flows to zero.  We will focus on the behavior for $x \ll 1$, near the phase transition.

If we consider systems with quenched randomness then the decay rate will vary from one insulating region to another.  Here we will first ignore these variations, making the approximation of a spatially uniform $x$.  Below we will revisit the question of whether spatial variations in $x$ are important, after the critical behavior is determined, arguing that they are not relevant at the critical point.

The other RG parameters should relate to the density of T-blocks in space. In the RG at cutoff $\Lambda$, adjacent T-blocks combine if the insulating region between them is of length $\Lambda/x$. This process does not produce correlations between T-blocks other than the implied ``hard-core'' condition of a minimum spacing between T-blocks.  Thus if we start with a bare system with quenched randomness and only short-range correlations in the local properties, at large cutoff $\Lambda$ it will produce a distribution of spacings between T-blocks that is exponential, beyond the minimum-distance condition. This means that it makes sense to define $R_\Lambda$ as the rate (in space) of T-block occurrence along the line when the cutoff is $\Lambda$, so that $R_\Lambda \exp(-R_\Lambda w)dw $ is the probability that the insulating region between two adjacent T-blocks has its
%left endpoint of the first T-block is after an interval (I-block) with
length in the interval $ [\Lambda/x + w, \Lambda/x + w+dw]$. Note that $R_\Lambda$ does not give the density of T-blocks, since it is the rate of their occurrence {\it outside of} the minimum-distance constraint, and excluding their own length. This overall rate $R_\Lambda$ may be broken down into separate rates for each possible length $\ell \ge \Lambda$ of the next T-block along the line:
\begin{equation}\label{2.1}
R_\Lambda = \int_\Lambda^\infty r_\Lambda(\ell) d\ell.
\end{equation}
Thus $r_\Lambda(\ell)d\ell / R_\Lambda$ is the probability that the next T-block has length $\in [ \ell,  \ell+d\ell]$, and $r_\Lambda(\ell)d\ell$ is the rate of occurrence of T-blocks in that length range.

%\am{Added this. I think some discussion like this should go somewhere. Here seemed okay.}At large lengthscales when the RG is in the regime of universality, larger T-blocks will occur at lower rates at a fixed RG cutoff. Thus $r_\Lambda(\ell)$ should be a decreasing function of $\ell$ at fixed $\Lambda$. T-blocks of length $\ell$ are generated when the RG scale is $\lessapprox x \ell$ (and more so as this upper limit is approached), so as a function of $\Lambda$ we expect $r_\Lambda(\ell)$ at fixed $\ell$ increases until the cutoff reaches $x\ell$. After that (at larger cutoffs), the only processes involving T-blocks of that size $\ell$ are ones that deplete those T-blocks by combining them to make larger ones. Thus $r_\Lambda(\ell)$ should be decaying for $\Lambda \in [x\ell,\ell]$, before abruptly going to zero as the cutoff $\Lambda$ reaches $\ell$ and all the remaining T-blocks of that size are absorbed into insulating regions. Notably, in regimes in which the localization of T-blocks is far more common than the creation of new, large T-blocks, most T-blocks get erased at $\Lambda=\ell$ rather than contributing to the creation of larger T-blocks before that, and so $r_\Lambda(\ell)$ should only be weakly decaying when $\Lambda\in [x\ell, \ell]$. This is the case in the critical and MBL regimes of interest that we consider in this work.

Let us now focus on $r_\Lambda(\Lambda)$. This is the rate for T-blocks whose length is at the cutoff scale $\Lambda$. Having failed to combine into larger T-blocks, all remaining T-blocks at the cutoff scale are  erased as $\Lambda \rightarrow \Lambda + d\Lambda$. As $r_\Lambda(\Lambda)$ has dimensions $1/(\text{length})^2$,  we may define a dimensionless rate $y = y_\Lambda = \Lambda^2 r_\Lambda(\Lambda)$. We will use $y$  as the second RG parameter when we consider two-parameter RG flows or recursions near the phase transition. (Note that $y$ does depend on $\Lambda$, but usually we will leave off the subscript.) As we will see, these two parameters provide a  useful way of encapsulating the behavior of the functional RG for the flow of $r_\Lambda(\ell)$ and  $x=x_\Lambda$ with $\Lambda$. This works well, provided $x$ and $y/x$ are small, which covers the critical region up to the point at which the RG flows strongly towards thermalization.

The MBL phase is governed by a fixed line with $r_\Lambda(\ell)=0$ and thus $y=0$, which means the system is asymptotically all insulating with no T-blocks.  This MBL fixed line is parametrized by $x\ge 0$ which is set by the decay length of the interactions in this insulator.  For $x>0$ this fixed line is stable to adding a low enough density of T-blocks.  It becomes unstable at $x=0$, so the MBL phase transition is 
%We expect that the critical behavior should be 
governed by the fixed point $x=y=0$ at the terminus of the MBL fixed line. 

For small $x$ and $y$, the dominant mode of production of T-blocks of size $\Lambda/x$ should be the combination of component T-blocks of size close to $\Lambda$. This is based on the idea that the most efficient (\textit{i.e.} least unlikely) way to create a T-block of a given length is to combine the smallest possible sub-blocks. In this approximation, we should have
\begin{equation}\label{2.2}
r_\Lambda(\Lambda/x) = R_\Lambda^2.
\end{equation}
For the same reason, $r_\Lambda(\ell)$ should be weakly dependent on $\Lambda$ between $x\ell$ and $\ell$, as any such dependence involves combination of  sub-blocks that are larger than the minimum size. Thus $r_\Lambda(\ell) \approx y_{\ell}/\ell^2$ for $\Lambda \le \ell\le \Lambda/x$. We would like to approximate $R_\Lambda \approx \Lambda r_\Lambda(\Lambda)$. This is valid if $y$ is a slowly varying function of $\Lambda$, since then $R_\Lambda$ can be approximated by $\int_\Lambda^\infty (y_\Lambda/\ell^2) d\ell = y_\Lambda/\Lambda = \Lambda r_\Lambda(\Lambda)$. It will turn out that $y$ is indeed slowly varying on the separatrix of the flow.  Near the fixed line, we will see that $y$ is not slowly varying, but nevertheless the approximation is not so bad as to spoil our conclusions about the behavior there. (In Sec.~\ref{sec:more_accurate}, we will show how to improve on this.)

If we insert $R_\Lambda = \Lambda r_\Lambda(\Lambda)$ in Eq.~(\ref{2.2}), we obtain $r_\Lambda(\Lambda/x) = \Lambda^2 r_\Lambda(\Lambda)^2$. From the weak dependence of $r_\Lambda$ on $\Lambda$, we have $r_\Lambda(\Lambda/x) \approx r_{\Lambda/x}(\Lambda/x)$, and so we obtain a recursion for $y$:
\begin{equation}\label{2.3}
y_{\Lambda/x} = \left( \frac{y_\Lambda}{x_\Lambda} \right)^2.
\end{equation}

To complete the picture, we need a flow equation for $x$. As explained above, the T-blocks with size $\in [\Lambda, \Lambda+d\Lambda]$ are erased with the increment $\Lambda \rightarrow \Lambda + d\Lambda$. Since interactions do not decay across these regions of size $\Lambda$, and they occur at rate $r_\Lambda (\Lambda) d\Lambda$, the decay rate of interactions, $1+x$, gets reduced by $-\Lambda r_\Lambda (\Lambda) d\Lambda$ due to this increment. 
%For this, it is preferable to write $\Lambda = e^t$, where $t$ is the ``RG time". As explained above, the T-blocks with size $\in [\Lambda, \Lambda e^{dt}]$ are erased with the increment $t\rightarrow t+dt$. Two factors of $\Lambda$ (one from the size of these T-blocks, which gives the spatial extent of halted decay, and one from the change of variable) convert $r_\Lambda(\Lambda)$ to $y$, and we obtain the flow of $x$:
Changing to ``RG time" $t=\log\Lambda$ and converting the resulting factor of $\Lambda^2 r_\Lambda (\Lambda)$ to $y$, we obtain the flow of $x$:
\begin{equation}\label{2.4}
\frac{dx}{dt} = -y,
\end{equation}
otherwise known as the rule of halted decay~\cite{Thiery-DeRoeck2017b}. One of the motivations for our definition of $y$ is so that Eq. (\ref{2.4}) would be true for small $x$, $y$.

We have seen that the recursion/flow given by Eqs.~(\ref{2.3}) and (\ref{2.4}) follows from some general arguments based on the dominant  mode of creation of T-blocks of a given length and an assumption of uniform $x$. Analysis of the recursion will provide a quick route to an understanding of the behavior of the full RG for the rate function $r_\Lambda(\ell)$.

\section{Critical behavior\label{sec:critical}}

It should be clear from Eq.~(\ref{2.3}) that the separatrix is asymptotic to the curve $y = x^{2}$. If we start with a point on the curve $y = x^{2+\delta}$, then for small enough $x$, the image is essentially on the curve $y = x^{2+2\delta}$ (the change in $x$ from $dx/dt = -y$ being negligible over a $\Delta t = \log x^{-1})$.

To understand the flow along the separatrix, observe that substituting $y=x^2$ in Eq.~(\ref{2.4}) leads to
$dx/dt = -x^{2},$
which has  solution
$x = t^{-1}$, and hence $y \approx x^{2} = t^{-2}$.
Recall that $t = \log \Lambda$. Note that $y \approx 1/(\log \Lambda)^2$ is indeed a slowly varying function of $\Lambda$; this verifies one of the assumptions made in deriving Eq.~(\ref{2.3}).

In order to examine the rate of departure from the separatrix during the RG recursion/flow, we may consider as above the evolution of $\delta = 
%\frac{\log y}{\log x} -2
(\log y)/ (\log x) - 2$.
We have, in essence, an exponential rate of departure from the separatrix with respect to the number, $n$, of recursion steps. So if we start at a small departure $\delta_0$, then $n = \log_2 \delta_0^{-1}$ steps are required until $\delta = O(1)$. Let $T$ denote the RG time required for those $n$ steps. Progress along the separatrix has a logarithmic slowdown, \textit{i.e.}, RG time steps in the recursion are of size $\log x^{-1}\approx\log t$ near the separatrix, so 
%the majority of the steps are of size  $\sim \log T$, 
$\Delta T / \Delta n = \log T$,
and hence $n \approx \int_2^T dt/\log t = \text{Li}(T) 
%\frac{\log \delta_0^{-1}}{\log 2}
%\log \delta_0^{-1} / \log 2
\approx T/\log T$. Thus $T \approx 
%\frac{\log \delta_0^{-1}}{\log 2}
%( \log_2 \delta_0^{-1} ) \log \log \delta_0^{-1}$, 
n\log n + n\log \log T$, and dropping the subdominant second term, we obtain 
$T \approx (\log_2 \delta_0^{-1} )\log \log_2 \delta_0^{-1}$, which leads to
 a diverging length
\begin{equation}\label{3.1}
\Lambda = e^T = \delta_0^{-\log_2 \log_2 \delta_0^{-1}}.
\end{equation}
This evidently diverges faster than any power of $\delta_0$, so we have in effect $\nu = \infty$. 
Thus 
the Chayes, \textit{et al.} inequality $\nu \ge 2$ is satisfied~\cite{Harris1974,Chayes-Spencer1986,Chandran-Oganesyan2015}. In essence, this is telling us that fluctuations of size $\Lambda^{-1/2}$ in $x$ are small in comparison to the initial displacement $\delta_0 = \Lambda^{-1/\nu}$ that is needed to depart the vicinity of the separatrix at scale $\Lambda$. In this way, we can justify the neglect of fluctuations in $x$ in the derivation of these equations.

It is noteworthy that in a finite-size system where $\Lambda$ cannot be larger than the physical size of the system, we can obtain an expression for the effective system-size-dependent $\nu(\Lambda)$: First we identify 
$\nu(\Lambda) = (\log \Lambda)/(\log \delta_0^{-1}) = T/(n\log 2)$, then using $n\approx T/\log T$, this becomes $\nu(\Lambda) \approx \log_2 T = \log_2 \log \Lambda$.
%$\nu(\delta_0) =  \log_2 \log \delta_0^{-1}$ from Eq.~(\ref{3.1}), then we use Eq.~(\ref{3.1}) to eliminate $\delta_0$ in favor of $\Lambda$. The result is that $\nu(\Lambda)$ is the solution to $\log(\Lambda)  =\nu 2^\nu$. 
This implies that $\nu(\Lambda)$ is an extremely slowly increasing function of $\Lambda$. 
For example, $\nu(10^6) \approx 3$ (here we use $\text{Li}(T)$ for $n$ instead of $T/\log T$).
%Numerically solving for $\nu(\Lambda)$ we see that, for example, in order for $\nu > 4$ it must be true that $\Lambda > 10^{27}$. 
Therefore, direct finite-size numerical approaches will not be able to access the asymptotic criticality that we study in this work. But $\nu(\Lambda)$ is consistent with previous numerical studies of the MBL transition in approximate RG models, in which $\nu \approx 3$ was consistently found~\cite{Vosk-Altman2015,Potter-Parameswaran2015,Dumitrescu-Potter2017,Thiery-DeRoeck2017a}.
%Nonetheless it is worth noting that $\nu(\Lambda)$ varies from 2.1 to 2.5 for $\Lambda$ between $10^4$ and $10^6$; this is consistent with the numerics in Ref. \cite{Morningstar-Huse2019}.
%\jzi{Do you want to cite a couple of those studies then?}

For comparison, the usual KT flow also exhibits logarithmic slowdown and $\nu = \infty$; however in that case progress is slow both along the separatrix and orthogonal to it. Here we still have what is essentially exponential divergence from the separatrix, but it proceeds through the logarithmically-slowed RG time that is dictated by the separatrix flow. Consequently the form of the divergence of $\Lambda$ as $\delta_0 \rightarrow 0$ given by Eq.~(\ref{3.1}) is very different from the standard KT picture.

This unusual combination of slow and fast modes leads to another important feature of this universality class: extreme narrowness of the critical window. This can be quantified by $x_f$, the 
% initial displacement $\delta_0>0$ that leads to a given 
final value of $x$ on the MBL fixed line that follows from an initial displacement $\delta_0>0$. Since the flow of $x$ effectively freezes once $\delta = O(1)$ (the system is decidedly in the MBL phase), and $x(t) \approx 1/t$ on the separatrix, we can infer from Eq.~(\ref{3.1}) that $x_f$ is approximately given by 
$1/T= [(\log \delta_0^{-1}) \log_2 \log_2 \delta_0^{-1}]^{-1}$. Consequently, the window width is given by $\delta_0(x_f) \approx \exp[-1/(x_f \log_2 x_f^{-1})]$; compare with $\delta_0 \approx x_f^2$ for the standard KT flow. This indicates how finely-tuned the initial condition needs to be to keep the system critical down to $x\sim x_f$.

% While the critical behavior can be seen directly from Eqs.~(\ref{2.3}) and (\ref{2.4}), it is instructive to derive an equivalent continuous flow. The natural way to do this is to replace Eq.~(\ref{2.3}) with
% \begin{equation}\label{3.2}
% \frac{dy}{dt} = -(\log 2)y \delta = -(\log 2)y \left(\frac{\log y}{\log x} -2\right).
% \end{equation}
% From this, we see that the separatrix is still asymptotic to the curve $y=x^2$, and $x \sim 1/t$, $y\sim 1/t^2$ still holds. We compute that $d\delta/dt \approx 
% %\frac{\log 2}{\log x^{-1}}
% \delta(\log 2) / (\log x^{-1}) $ (as before, the contribution from the flow of $x$ is negligible). As in the case of the recursion, we have that $\delta \rightarrow 2\delta$ when $t \rightarrow t+\log x^{-1}$. From this, we may derive the same behavior of the diverging length shown in Eq.~(\ref{3.1}) as one approaches the MBL transition.
% See Fig.~\ref{fig:2d_flow} for a stream plot of the RG flow given by Eqs.~(\ref{2.4}) and (\ref{3.2}).

While the critical behavior can be seen directly from Eqs.~(\ref{2.3}) and (\ref{2.4}), it is instructive to derive an equivalent continuous flow to replace the discrete recursion for $y$. We began this section by noting that near the separatrix Eq.~(\ref{2.3}) is equivalent to the recursion $\delta \to 2 \delta$ when $t \to t+ \log x^{-1}$, where $\delta = (\log y) / (\log x) - 2$ and $t = \log \Lambda $. Again treating the contribution from the flow of $x$ as negligible, this discrete recursion is reproduced by the continuous equation 
\begin{equation}\label{delta}
\frac{d \delta}{d t} \approx \frac{\log 2}{\log  x^{-1}  }  \delta.
\end{equation}
From this we may derive the same behavior of the diverging length shown in Eq.~(\ref{3.1}) as one approaches the MBL transition because this continuous flow reproduces the discrete recursion. Translating back to an equation for $y$ we arrive at
\begin{equation}\label{3.2}
\frac{dy}{dt} = -(\log 2)y \delta = -(\log 2)y \left(\frac{\log y}{\log x} -2\right).
\end{equation}
From Eq.~(\ref{3.2}) we can see that the separatrix is still asymptotic to the curve $y=x^2$, and $x \sim 1/t$, $y\sim 1/t^2$ still holds. See Fig.~\ref{fig:2d_flow} for a stream plot of the RG flow given by Eqs.~(\ref{2.4}) and (\ref{3.2}).

It is natural to ask what is the origin of the unusual logarithmic terms in the flow equation. With careful definitions for RG parameters $x$, $y$, we were able to derive a very simple RG recursion relation for these parameters, Eq.~(\ref{2.3}). This equation involves the ratio $y/x$ and a squaring operation. These operations only make sense at an infinitesimal level if one works with the logarithms of $x$, $y$. In fact the flow equation (\ref{delta}) for $\delta = (\log y)/(\log x) -2$ is linear in $\delta$ (although the rate of change is reduced by a factor of $\log x^{-1}$). This makes it clear that $\delta = 0$ defines the separatrix, and the rate of departure from it is proportional to $1/(\log x^{-1})$.  This leads to the form (\ref{3.1}) of the divergence of the length scale for departure from criticality. Although this flow agrees with traditional KT flow (and with \cite{Goremykina-Serbyn2018,Dumitrescu-Vasseur2018}) on the separatrix, the discussion above shows that the behavior elsewhere (in particular the rate of departure from the separatrix) is quite different.

Near the $x$-axis, well below the separatrix, we see that for both discrete and continuous equations,  $y$ exhibits super-exponential decay, and so $x$ effectively freezes.  In terms of $z = \log y$ Eq.~(\ref{2.3}) becomes 
\begin{equation}\label{3.2a}
z(\Lambda/x) = 2z(\Lambda) - 2 \log x,
\end{equation}
 and so for $\left|z \right| \gg \left| \log x \right|$, we have that
\begin{equation}\label{3.3}
z(\Lambda) \approx -2^{\log_{x^{-1}} \Lambda} = -2^{\log \Lambda/\log x^{-1}} = -\Lambda^{\log 2/\log x^{-1}}.
\end{equation}
This corresponds to $y = \exp(-\Lambda^\xi)$, with $\xi$, the fractal dimension, equal to $(\log 2)/(\log x^{-1})$. This agrees with the result of Ref.~\cite{Morningstar-Huse2019} in the limit $x \ll 1$ that we have been considering. Note that $\xi$ tends to zero as one tends towards the transition at $x=0$.

The continuous equation has identical behavior near the $x$-axis. Rewriting Eq.~(\ref{3.2})  in terms of $z$, we have that
\begin{equation}\label{3.4}
\frac{dz}{dt} = (\log 2)\left(\frac{z}{\log x^{-1}} -2\right).
\end{equation}
For large negative $z$, we may neglect the 2 in Eq.~(\ref{3.4}), and the resulting exponential growth reproduces Eq.~(\ref{3.3}) after replacing $t$ with $\log \Lambda$.

Above the separatrix, it is evident that once $\delta$ is $O(1)$, both the recursion and the flow leave, in finite RG time, the regime of their validity (that is, $x$ and $y/x$ small). We presume, then, that within a finite RG time, the majority of space will be covered by T-blocks, and the system is decidedly approaching complete thermalization.

Recall that $r_\Lambda(\ell) \approx r_\ell(\ell)$ for $\Lambda \le \ell\le \Lambda/x$ (see also Eq.~(\ref{independence}) below). This means that any solution $(x(t),y(t))$ to the flow determines $r_\ell(\ell) = y_\ell/\ell^2$ as the (unnormalized) distribution of T-block sizes in $[\Lambda,\Lambda/x]$ when the cutoff is $\Lambda$. We assumed from the beginning that this distribution is dominated by $\ell$ near $\Lambda$, and this is evidently true on the separatrix (where $y_\ell \sim 1/{(\log \ell)^2}$) and below (where $y_\ell$ decreases more rapidly). We see that the critical theory exhibits a $1/\ell^2$ distribution, with a logarithmic correction. This is consistent with all of the previous RGs, which found a distribution of T-block sizes approaching a power law $\propto \ell^{-\alpha}$ at criticality, with $\alpha \gtrsim 2$~\cite{Goremykina-Serbyn2018,Dumitrescu-Vasseur2018,Morningstar-Huse2019}. Noting that $x_\Lambda^{-1} \approx t = \log \Lambda$, we see that the average size of T-blocks for the critical theory at cutoff $\Lambda$ is approximately
\begin{align}\label{size}
R_\Lambda^{-1}\int_\Lambda^{\Lambda/x}\ell r_\ell(\ell) d\ell &= 
R_\Lambda^{-1}\int_\Lambda^{\Lambda/x} \frac{ d\ell}{\ell(\log \ell)^2} \\ \nonumber
&\approx R_\Lambda^{-1}\frac{\log x^{-1}}{(\log \Lambda)^2} \approx R_\Lambda^{-1}\frac{\log \log \Lambda}{(\log \Lambda)^2}.
\end{align}
Recalling that  $R_\Lambda \approx y_\Lambda/\Lambda$, the average size of I-blocks is $\Lambda/x + R_\Lambda^{-1}= R_\Lambda^{-1}(y/x + 1) \approx R_\Lambda^{-1}$. Thus for the critical theory the fraction of the system in T-blocks decreases as $(\log \log \Lambda)/(\log \Lambda)^2$.

\section{A concrete RG and its flow equations\label{sec:concrete_RG}}

In this section we introduce the RG of Ref.~\cite{Morningstar-Huse2019}, which was, in turn, a modification of the RGs of Refs.~\cite{Goremykina-Serbyn2018,Zhang-Huse2016}, and modify it so as to work in the approximation of spatially uniform $x$ within the insulating regions. The resulting flow equations for $x$ and $r_\Lambda(\ell)$ can be written down exactly. We examine these under the assumption that $x$ and $y/x$ are small, and show that our fundamental recursion relation Eq.~(\ref{2.2}) follows. For definiteness, let us assume that $y \le x^{3/2}$.

Following Ref.~\cite{Morningstar-Huse2019}, the line is divided into a sequence of alternating T-blocks (thermalized blocks) and I-blocks (insulating blocks). At a given RG cutoff length scale $\Lambda$, the T-blocks have lengths $\ell \ge \Lambda$. The I-blocks are characterized by two lengths, the physical length $\ell$ and the ``deficit'' $d$. The latter can be interpreted as the length of the shortest T-block that can, by itself, thermalize that I-block. At this point the parameter $x$, which describes how close an I-block is to the avalanche instability~\cite{Thiery-DeRoeck2017b,DeRoeck-Huveneers2017,Luitz-DeRoeck2017,DeRoeck-Imbrie2017}, is given by $x = d/\ell$ and varies from one I-block to another. When the cutoff is $\Lambda$, all I-blocks have deficit $d \ge \Lambda$ and physical length $\ell \ge \Lambda/x$.
%; their physical length $L= d/x$ can be much larger. 
%We call the lengths $L$ and $d$ of T and I-blocks, respectively, the ``primary lengths" of the blocks. 
As the cutoff is raised from $\Lambda$ to $\Lambda + d\Lambda$, all T-blocks with $\ell$ and I-blocks with $d$ in that range are ``erased'' or absorbed, along with the two adjacent blocks, into a single new block whose physical length is the sum of the individual physical lengths.
%, and whose character (T or I) is the same as that of the adjacent blocks. More simply, the moves are 
These ``moves'' are either TIT$\rightarrow$T or ITI$\rightarrow$I. In the latter case, one sets $d_{\text{new}} = d_1 - \Lambda + d_2$, where $d_1$ and $d_2$ are the deficits of the two I-blocks.

From this starting point, we modify the RG to have the same $x$ across all I-blocks, or equivalently, the same decay length $\zeta$.
%, and we choose units where $\zeta_{\text{c}}= 1$ is the critical ``avalanche''threshold. Then $x$ corresponds to $\zeta^{-1} - 1$, and $d = xL$ for all I-blocks. 
The order of moves is as described above: when the cutoff length is $\Lambda$, TIT$\rightarrow$T moves happen when the middle block has $d =\Lambda$
(\textit{i.e.}, $\ell =\Lambda/x$), and ITI$\rightarrow$I moves happen when the middle block has $\ell =\Lambda$. The TIT$\rightarrow$T moves do not change the global $x$, since
they do not make new I-blocks, but the ITI$\rightarrow$I moves do. When an ITI$\to$I move happens, the new I-block is first generated as defined above. But that I-block then has a new value of $d/\ell$ that is different from the global value of $x$, so we ``average" over all I-blocks to compute a new global $x$ and use that to reset the deficit $d$ of all I-blocks to $d=x\ell$. This ensures the total length of the system is preserved. When the RG length cutoff is  $\Lambda$, TIT$\to$T
moves generate T-blocks of size $> (2+x^{-1})\Lambda \approx\Lambda/x$ and the ITI$\to$I moves generate
I-blocks of size $ > (2x^{-1} + 1) \Lambda \approx 2\Lambda/x$. Both types of moves are capturing processes
at physical time $\exp(c\Lambda/x)$ for some
 order-one constant $c$, because they are both
associated with an avalanche running for a distance $\Lambda/x$ (either across the I-block as an I-block thermalizes or into I-blocks as a T-block localizes). Interblock correlations are
not generated by these RG rules because the order of moves is determined only by the
properties of the middle blocks in any candidate move.

In the context of this RG, one may define as in Sec.~\ref{sec:recursion} the rate functions $r_\Lambda(\ell)$ and $R_\Lambda = \int_\Lambda^\infty r_\Lambda(\ell)d\ell$. In terms of these quantities, the exact flow equations are as follows:
\begin{widetext}
\begin{align}\label{4.1}
\frac{dx}{d\Lambda} &= - \frac{\Lambda r_\Lambda(\Lambda)(1+x)}{1+\Lambda R_\Lambda/x} \\
\label{4.2}
\frac{dr_\Lambda(L)}{d\Lambda} &= \frac{1}{x}\left(\frac{dx}{d\Lambda} -R_\Lambda\right) r_\Lambda(L) + \frac{1}{x}\Theta(L-[2+x^{-1}]\Lambda) \int_\Lambda^{L-(1+x^{-1})\Lambda} d\ell r_\Lambda(\ell)r_\Lambda(L-\ell -\Lambda/x).
\end{align}
\end{widetext}
See Appendix~\ref{app:RG_eqns} for details. 

Recall that $y \equiv \Lambda^2 r_\Lambda(\Lambda)$ is the dimensionless version of the rate at the cutoff length, $\ell = \Lambda$. As in Sec.~\ref{sec:recursion}, it may be convenient to switch out $R_\Lambda$ for $\Lambda r_\Lambda(\Lambda)$. As these two quantities behave similarly, let us define a ``correction factor'' $a = \Lambda r_\Lambda(\Lambda) / R_\Lambda$. Then $\Lambda R_\Lambda = \Lambda^2 r_\Lambda(\Lambda)/a =y/a$. If we insert this into Eq.~(\ref{4.1}) and switch to $t=\log \Lambda$, we obtain 
\begin{equation}\label{4.3}
\frac{dx}{dt} = -\frac{y(1+x)}{1+ y/(ax)}.
\end{equation}
This makes it clear that for $x$ and $y/x$ small, Eq.~(\ref{4.3}) becomes $dx/dt = -y$, as claimed earlier in Eq.~(\ref{2.4}). (We will see in the next section that $a$ stays away from 0 as long as $x$ and $y/x$ are small; also  $a \approx 1$ along the separatrix.)

The main contribution in Eq.~(\ref{4.2}) should be the final term, which represents TIT$\to$T moves producing new T-blocks of length $L$. The first term gives small contributions from implicit dependence on $x$ in $r_\Lambda(L)$, and from T-blocks of length $L$ absorbed into larger T-blocks in TIT$\to$T moves.

We may obtain an expression for $r_\Lambda(L)$ at $L = (2+x^{-1})\Lambda$ by integrating Eq.~(\ref{4.2}) $d\Lambda^\prime$ from $x\Lambda$ to $\Lambda$, say. The lower limit does not matter much because for such $L$ the initial condition is negligible. 
(In general, we expect that $r_\Lambda(\ell)$ should decay exponentially for $\ell > \Lambda/x$, with a decay length $\le \Lambda/x$. This is based on the principle that the least-unlikely way to produce a T-block with a given length is to join up T-blocks near the cutoff scale $\Lambda$, separated by I-blocks of size $\Lambda/x$. In the case at hand, $r_{x\Lambda} (\ell)$ should have a decay length  $\le \Lambda$, so $r_{x \Lambda} (L) $ is suppressed by a factor $e^{-1/x}$.)
%(The initial $r_{\Lambda/4}(L)$ would be based on merging a minimum of three T-blocks, so it would be down by at least one factor of $y$, compared with $r_{\Lambda}(L)$.)
%\am{I don't follow this three T-block statement. The way I see it, to generate a new T-block of size $\sim \Lambda/x$ when the cutoff is $\Lambda/4$ requires two T-blocks that add up to $\sim 3 \Lambda / 4 x$, and this is down by at least powers of $x$, so generally this integral is dominated near its upper limit (near $\Lambda'>\alpha \Lambda$ for $1-\alpha \ll 1$).} 
%\jzi{Here we're getting into the reason for rapid falloff of $r_\Lambda(L)$ for $L > \Lambda/x$. I'm relying here on the idea that the ``least unlikely'' way to span $\Lambda/x$ when the cutoff is $\Lambda/4$ would be with three T-blocks. If you use two, one of those is big and it would have to be formed from another TIT move. Basically, most T-blocks should have size comparable to the cutoff. Maybe $\Lambda/4$ is cutting it too fine. I could just as well replace that with $x\Lambda$ (it would just add a log) and then it should be a little more obvious that the lower-limit term is negligible. Do you think that would be preferable?} 
Therefore, let us ignore the initial condition and write
\begin{align}
r_\Lambda(L) =& \int_{x\Lambda}^\Lambda d\Lambda^{\prime} \bigg[ \frac{1}{x}\left(-\frac{y}{\Lambda^\prime} -\frac{y}{a\Lambda^\prime}\right) r_{\Lambda^\prime}(L)+ \nonumber \\
\label{4.4}
& \frac{1}{x} \int_{\Lambda^\prime}^{L-(1+x^{-1})\Lambda^\prime} d\ell
r_{\Lambda^{\prime}}(\ell)r_{\Lambda^{\prime}}(L-\ell-\Lambda^\prime/x) \bigg].
\end{align}

Here we used $dx/d\Lambda = (1/\Lambda)dx/dt = -y/\Lambda$ and switched out $R_\Lambda$ for $y$ and $a$ as described above. The $\Theta$-function is 1 for the $L$ we are considering ($L=[2+x^{-1}]\Lambda$). We claim the second term is approximately $R_\Lambda^2 = [y/(a\Lambda)]^2$, and the first term is much smaller than this. 

In the second term, we change variables, letting $\ell^\prime = L - \ell - \Lambda^\prime/x$. Then $d\Lambda^\prime = xd\ell^\prime$, which cancels the factor $1/x$, leaving the following:
\begin{equation}\label{4.5}
\int d\ell^{\prime} d\ell
r_{\Lambda^{\prime}}(\ell)r_{\Lambda^{\prime}}(\ell^\prime).
\end{equation}
Here $\Lambda^\prime = x(L-\ell-\ell^\prime)=\Lambda +x(2\Lambda - \ell - \ell^\prime)$ is now a function of $\ell$, $\ell^\prime$. This looks a lot like $R_\Lambda^2$, 
because at least on the domain $\{\ell$, $\ell^\prime \ge \Lambda\}$, weak dependence of $r_\Lambda(\ell)$ on $\Lambda$---see Eq.~(\ref{independence}) below---allows us to replace $\Lambda^\prime$ with $\Lambda$ in Eq.~(\ref{4.5}). 
For the same reason, $r_\Lambda(\ell) \approx r_\ell(\ell) = y_\ell/\ell^2$, so from the $1/\ell^2$ decay and/or from the decay of $y_\ell$, it is clear  that Eq.~(\ref{4.5}) is dominated by $\ell$, $\ell^\prime$ near $\Lambda$ (and consequently $\lvert\Lambda^\prime - \Lambda\rvert \sim x\Lambda)$.
Understanding the full domain for $\ell$, $\ell^\prime$ is a little more complicated, but we show in Appendix~\ref{app:approximation} that Eq.~(\ref{4.5}) is equal to $R_\Lambda^2$, up to terms that are negligible for $x$ and $y/x$ small.  
This shows in detail how T-blocks of size $\Lambda/x$ are mainly formed from sub-blocks of size around $\Lambda$, as claimed in the heuristic argument for Eq.~(\ref{2.2}).

Let us return to the first term in Eq.~(\ref{4.4}). It represents the depletion of T-blocks that have previously been produced by the second term. For a crude estimate, we can apply this decay to the entire second term (even though, as we have seen, the second term is largely produced by $\Lambda^\prime$ in a window of size $\sim x\Lambda$ in front of $\Lambda$). Writing $s = \log \Lambda^\prime$, $d\Lambda^\prime/\Lambda^\prime = ds$, we see that the decay occurs over a range of $s$ of size $\log x^{-1}$, so the decay factor is $\exp \left[-\int ds (y/x)\right] \approx 1-O(y/x)\log x^{-1}$. Thus the first term is bounded by $O[(y/x) \log x^{-1}] R_\Lambda^2$; recall that by assumption $y/x \le x^{1/2}$.

A similar analysis may be performed to determine the fate of the T-blocks after their production at $\Lambda^\prime \approx \Lambda$. Note that from Eq.~(\ref{4.2}), it is clear that for $L = (2+x^{-1})\Lambda$ (as we have been considering) the $\Theta$-function implies that there is no further production for $\Lambda^\prime > \Lambda$. Thus we may obtain $r_L(L)$ by again applying the decay over a range of $s$ of size $\log x^{-1}$, which again results in a drop of size $O[(y/x) \log x^{-1}] R_\Lambda^2$, so $r_\Lambda(L)$  is almost independent of $\Lambda$ for $\Lambda \in [xL,L]$. Thus
\begin{equation}\label{independence}
r_\Lambda(L)  = r_L(L)(1-O[(y/x) \log x^{-1}]).
\end{equation}
%(Recall that $\Lambda$ cannot exceed $L$, because when it reaches $L$, all remaining T-blocks of length $L$ are eliminated in ITI$\to$I moves.) 
This statement gives a quantitative version of the approximation $r_\Lambda(\Lambda/x) \approx r_{\Lambda/x}(\Lambda/x)$, which was used to convert Eq.~(\ref{2.2}) into the key recursion Eq.~(\ref{2.3}) for $y_\Lambda \equiv \Lambda^2 r_\Lambda(\Lambda)$.

In conclusion, we have the approximate relationship $r_\Lambda([2+x^{-1}]\Lambda) = R_\Lambda^2$, and then to the level of approximation that we have been working with, this can be replaced with
\begin{equation}
r_\Lambda(\Lambda/x) = R_\Lambda^2,
\end{equation}
which is Eq.~(\ref{2.2}). We may switch out $R_\Lambda = \Lambda r_\Lambda(\Lambda)/a$ and  $r_\Lambda(\Lambda/x) \approx r_{\Lambda/x}(\Lambda/x)$, to turn this into
\begin{equation}
r_{\Lambda/x}(\Lambda/x) = r_\Lambda(\Lambda)(\Lambda/a)^2.
\end{equation}
This may be expressed in terms of $y$:
\begin{equation}\label{4.10}
y_{\Lambda/x} = \left( \frac{y_\Lambda}{a_\Lambda x_\Lambda} \right)^2.
\end{equation}
However, this recursion still depends on the parameter $a$, which was set to 1 as a working hypothesis in Sec.~\ref{sec:recursion}.  The behavior of $a$ will be addressed in the next section.

\section{A more accurate flow\label{sec:more_accurate}}

In this section we will be using $t=\log \Lambda$ as the RG parameter, so
let us use $R(t)$ instead of the previous $R_\Lambda$. The rate function $r_\Lambda(\ell)$ is for T-blocks in the range $[\ell, \ell+d\ell]$, but in terms of $s = \log \ell$, the corresponding rate with respect to $ds$ is $\ell r_\Lambda(\ell)$. At the lower edge, this is $\Lambda r_\Lambda(\Lambda) = y/\Lambda$, and we will denote this by  $p(t)\equiv ye^{-t}$. We will use the approximate relation $R^\prime(t) \approx -p(t)$, where the prime denotes a derivative with respect to $t$. This is essentially the fundamental theorem of calculus [see Eq.~(\ref{2.1})], except that $r_\Lambda(\Lambda)$ depends on $\Lambda$ also in the subscript (the RG scale); however, as explained at the end of Sec.~\ref{sec:concrete_RG}, that dependence is weak, with the error down by a factor of $y/x$. 
(The dependence represents loss of weight in the distribution due to TIT moves---in the regimes we consider these are much rarer than ITI moves, which occur at rate $p(t)$.) Recall that $a$ was defined as $\Lambda r_\Lambda(\Lambda)/R_\Lambda$, so in the new variables it is written as $a(t) = p(t)/R(t) \approx -R^\prime(t)/R(t) = (-\log R)^\prime$. Thus in this approximation $a(t)$ is the instantaneous rate of exponential decay of $R(t)$.
%Equivalently, $1+a$  is the instantaneous power law for decay of $r_\Lambda(\ell)$ with $\ell$ at $\ell=\Lambda$.

Recall that in solving the recursion, we assumed that $a=1$ and found that on the separatrix, $x \sim t^{-1}$, $y \sim t^{-2}$, and hence 
 $p(t) = t^{-2}e^{-t}$. Its negative antiderivative is $p(t)[1-2/t + O(t^{-2})]$; and since $R^\prime = -p(t)[1+O(y/x)] = -p(t)[1+O(t^{-1})]$,  we should likewise have $R(t) = p(t)[1+O(t^{-1})]$.  Hence $a(t)= 1+O(t^{-1})$, which is consistent with our original choice for $a$. 

Near the $x$-axis, we found that $y(t)$ behaves like $\exp(-e^{\xi t})$, with $\xi = (\log 2)/(\log x^{-1})$. This is also the leading behavior of $R(t)$, so $a(t) \approx -R^\prime/R = \xi e^{\xi t}$.
Hence  $\log a$  is essentially $\xi t$, and then based on Eq.~(\ref{4.10}) the recursion for $z$ becomes
\begin{equation}\label{5.1}
z(t+\log x^{-1}) = 2z(t) + 2 \log x^{-1} -2\xi t.
\end{equation}
It should be clear that the rate of exponential growth of $z = \log y$ is still $\xi$, \textit{i.e.}, $\lim_{t\rightarrow\infty}t^{-1}\log(-z)= \xi$.

While the original assumption $a=1$ was sufficient to determine our main conclusions about the critical behavior, it will be useful to develop a systematic way of reintroducing  $a$ into the flow. The following flow equation accomplishes this goal:
\begin{equation}\label{5.2}
\frac{dy}{dt} =-(\log 2)y\left(\frac{\log y}{\log x} -2 - \frac{2\log a(x,y)}{\log x}\right).
\end{equation}
Here $a(x,y)$ is the solution to the following equation:
\begin{equation}\label{5.3}
a -1= (\log 2)\left(\frac{\log y}{\log x} -2 - \frac{2\log a}{\log x}\right).
\end{equation}
As the right-hand side involves only $\log a$, a recursive solution to Eq.~(\ref{5.3}) will converge very rapidly.

In order to obtain Eq.~(\ref{5.3}), let us  approximate $a$  as $-p^\prime/p$ instead of $-R^\prime/R$. This is reasonable, because we can expect that $(p^\prime/p)/(R^\prime/R) = RR^{\prime\prime}/{R^\prime}^2$ is close to 1. Thus we may put 
\begin{align}\label{a}
a(t) &\approx -p^\prime/p = (-\log p)^\prime \\
&= (-\log ye^{-t})^\prime = (-\log y)^\prime +1.\nonumber
\end{align}
As a check on the separatrix, put $R$ equal to the antiderivative of $p$, and then the ratio $RR^{\prime\prime}/{R^\prime}^2 = 1+O(t^{-2})$. (The approximation is good more generally for slowly-modulated exponential functions.)  As a further check on the approximation near the fixed line, one may put $R(t)$ equal to its leading behavior, $\exp(-e^{\xi t})$, and then the ratio $RR^{\prime\prime}/{R^\prime}^2 = 1-e^{-\xi t}$ is likewise close to 1, for  $t$ large. Within this approximation, we have that $a-1 = -y^\prime/y$. Then using Eq.~(\ref{5.2}) for $y^\prime/y$, we obtain Eq.~(\ref{5.3}), whose solution determines $a(x,y)$.  

To complete the picture, we need to derive Eq.~(\ref{5.2}). Recall that 
\begin{equation}\label{5.4}
\delta = \frac{\log y}{\log x} - 2.
\end{equation}
Observe that the uncorrected flow Eq.~(\ref{3.2}) can be obtained from the knowledge that in the recursion, $\delta \rightarrow 2\delta$ when $t \rightarrow t + \log x^{-1}$. Thus we may put $\delta^\prime = b\delta$ with $b= (\log 2)/(\log x^{-1})$.
This may be equated with the result from differentiating Eq.~(\ref{5.4}):
\begin{equation}\label{5.5}
\delta^\prime = \frac{y^\prime}{y} \cdot \frac{1}{\log x}.
\end{equation}
(Here we ignore the term $(-y\log y)/(x(\log x)^2)$ from differentiating $x$---compared with Eq.~(\ref{5.5}) it is down by a factor $y/x$, which is assumed to be small.)
We obtain the flow we have been working with:
\begin{equation}\label{5.6}
\frac{dy}{dt} =-(\log 2)y\delta=-(\log 2)y\left(\frac{\log y}{\log x} -2 \right).
\end{equation}

Now let us perform a similar procedure on the corrected recursion Eq.~(\ref{4.10}). In this case, when $t \rightarrow t+\log x^{-1}$, $\delta \rightarrow 2\delta - 2(\log a)/(\log x)$. This may be modeled by the ODE $\delta^\prime = b\delta + c$, whose solution  is
\begin{equation}
\delta(0)e^{bt} + \frac{c}{b}(e^{bt}-1).
\end{equation}
Now putting $t=\log x^{-1}$, and noting that $e^{bt} = 2$, we see that the recursion is satisfied if $c/b = - 2(\log a) /( \log x)$. Thus if we put $c = 2 [(\log 2) / (\log x)^2] \log a$, we obtain the desired value of $\delta(t)$.

To obtain the flow,  we may equate Eq.~(\ref{5.5})  with $b\delta +c$ to obtain
\begin{equation}\label{5.8}
y^\prime = y \log x\left(-\frac{\log 2}{\log x}\delta + c\right) 
%= -(\log 2)y\left(\delta - c\frac{\log x}{\log 2}\right)
=-(\log 2)y\left(\delta - \frac{2\log a}{\log x}\right),
\end{equation}
which is Eq.~(\ref{5.2}). This corrected equation can then be used to determine a better approximation for $y^\prime/y$ and hence $a$. Note that the $\log a$ term is much smaller than $(\log y) /( \log x) -2$ in all regimes except in a very small neighborhood of the separatrix. On the separatrix, we may put $y=t^{-2}$, and then $y^\prime/y = -2t^{-1}$. Consequently, Eq.~(\ref{5.6}) implies that $\delta = (2 / \log 2) t^{-1} =(2 / \log 2) x$, to leading order in $t^{-1}$ or $x$. 
(This leads to a corrected equation for the separatrix: $y = x^{2+\delta} = x^2\exp(2x\log_2 x) \approx x^2(1+2x\log_2 x)$, which as expected is slightly below the parabola $y = x^2$.)
As $a(t) = 1+ O(t^{-1})$, $\log a$ is $O(t^{-1})$, so the correction is still smaller than the rest,  by a factor of $\log x$. So practically speaking, one may replace Eq.~(\ref{5.3}) with
\begin{equation} \label{5.9}
a = 1+ (\log 2)\left(\frac{\log y}{\log x} -2 \right).
\end{equation}
On the separatrix, Eq.~(\ref{5.9}) gives a correction $O(t^{-1})$ to $a-1$, but this does not actually improve our understanding of $a$ there, because it adds to an error of equivalent size from the approximation $a\approx -R^\prime/R$. So Eq.~(\ref{5.8}) improves on Eq.~(\ref{5.6}) only away from the separatrix.

Note that $a$ is bounded away from 0 for $y \le x$, which verifies a claim made in Sec.~\ref{sec:concrete_RG}.
In fact, Eqs.~(\ref{5.2}) and (\ref{5.3}) show that if $a<1$, then $y^\prime > 0$, which implies that  $(x,y)$ is above the separatrix. 
%Recalling that $1+a$ is effectively the power law for the decay of $r_\Lambda(\ell)$ with $\ell$, we see that a power less than 2 occurs only when  the flow is on the way to thermalization.

If one runs the functional RG on an initial condition with a rapidly decaying distribution of T-block lengths, then one should expect that eventually the distribution will fill out to a particular value of $a$ satisfying Eq.~(\ref{5.3}). This procedure amounts to finding the stable manifold $a(x,y)$ in the three-parameter space $x$, $y$, $a$. This is possible because the flow for $x$, $y$ depends weakly on $a$, and because $a$ can be identified, at least approximately, from any candidate flow.

%But one should really do a more careful analysis to determine the asymptotic form of the separatrix.

%Let us return to the equation $\delta^\prime = b \delta$. If $b$ were $x$-independent, it would mean a fixed rate of departure from the separatrix, which would translate into a finite $\nu$. That is, starting from $\delta_0 \ll 1$, $\delta$ grows to $O(1)$ in time $t= \frac{1}{b}\log \delta_0^{-1}$, or length $\Lambda = \delta_0^{-1/b}$, hence $\nu = 1/b$. But since $b= \frac{\log 2}{\log x^{-1}}$, $\nu = \infty$. More precisely, we find that $\Lambda = \delta_0^{-(1/\log 2)\log \log \delta_0^{-1}}$, see (7.5) in my original notes. This analysis ties the unusual critical behavior very directly to the slowdown associated with lengthscale jumps of $1/x$, where  $x$ tends to zero as $1/t = 1/\log \Lambda$.

\section{Summary and discussion\label{sec:summary}}

In this work we analyzed the universal aspects of avalanche-driven MBL transitions in random one-dimensional systems. Building on previous works we employed a real-space RG approach and were able to understand the near-critical behavior analytically. 

Our results first followed from quite general arguments, and then using a slightly modified version of the RG of Ref.~\cite{Morningstar-Huse2019}, we were able to demonstrate these arguments within a particular RG using controlled approximations. This led into the explanation of how to systematically improve our approximations, but did not change the main results derived via more general considerations.

The resultant two-parameter RG flow was found to be %somewhat 
qualitatively similar to the Kosterlitz-Thouless (KT) flow, with the critical point being an unstable endpoint of the MBL fixed line; however some key differences were uncovered that put this MBL transition into a separate universality class. The lengthscale that diverges at the critical point does so more slowly (as a function of the bare displacement from criticality) in the case of the MBL transition than at KT transitions. This follows from a similarly slow flow along the separatrix, but a faster flow away from it. As a consequence, the critical window is extremely narrow, meaning that an exponential degree of tuning is needed in the bare system to ensure $x$ gets close to zero at large lengthscales.

This work contributes to an effort to understand the consequences of the avalanche mechanism for the MBL transition in one dimension. Due to the nature of the resulting criticality in an avalanche-driven MBL transition described above, it will be very difficult to verify our conclusions by numerically studying small systems. In fact, it has also proven difficult to study the asymptotic critical physics of the RGs numerically. Thus one promising route for progress is to continue to test the validity of the avalanche mechanism in more microscopic settings, theoretically, numerically, and experimentally.  

\begin{acknowledgments}

We thank Wojciech De Roeck, Romain Vasseur, and Sarang Gopalakrishnan for helpful discussions and comments on the manuscript. A.M. acknowledges the support of the Natural Sciences and Engineering Research Council of Canada (NSERC).  D.A.H. was supported in part by a Simons Fellowship and by DOE grant DE-SC0016244. J.Z.I. was supported by  Simons Foundation grant 638557.

\end{acknowledgments}

\appendix

\section{Derivation of functional RG equations\label{app:RG_eqns}}
In this appendix we derive Eqs.~(\ref{4.1}) and (\ref{4.2}), the flow equations for $x$ and $r_\Lambda (\ell)$ within the context of the RG described in Sec.~\ref{sec:concrete_RG}.

Let $n_\Lambda^{T}(\ell)$ be the number density (in block length, not in space) of T-blocks of length $\ell$ such that $N^T_\Lambda=\int_{\Lambda}^{\infty} n_{\Lambda}^{T}(\ell)d\ell$ is the total number of T-blocks in the system when the RG scale is $\Lambda$. This means that the probability density of a randomly chosen T-block having length $\ell$ is $p^T_\Lambda(\ell) = n^T_\Lambda(\ell) / N^T_\Lambda$. $n_{\Lambda}^{I}(d)$ is the corresponding number density for I-blocks, where we are using the deficit $d$ instead of the physical length, and the associated probability distribution over $d$ for I-blocks is $p^I_\Lambda(d) = n^I_\Lambda(d) / N^I_\Lambda$. Note that $N_{\Lambda}^{I}=N_{\Lambda}^{T}$ because blocks alternate between T and I, so we will drop the superscript.

When the cutoff goes from $\Lambda$ to $\Lambda+d\Lambda$, the number of TIT$\to$T moves that happen is $n^I_\Lambda(\Lambda)d\Lambda$, and the number of ITI$\to$I moves that happen is $n^T_\Lambda(\Lambda)d\Lambda$. The probability density (again, in block length) for the length $L$ of newly created T-blocks is
\begin{eqnarray}
\int_{\Lambda}^{\infty}d\ell p_{\Lambda}^{T}(\ell)p_{\Lambda}^{T}(L-\ell-x^{-1} \Lambda),
\end{eqnarray}
where $x^{-1}\Lambda$ is the contribution of the middle I-block to the length of the new T-blocks ($x$ is the same for all I-blocks in this model). Note that the upper limit $\infty$ on the integral can be moved down to $L-(1+x^{-1})\Lambda$ because $p_\Lambda^T(L-\ell - x^{-1}\Lambda) = 0$ for $\ell > L-(1+x^{-1})\Lambda$, but we will leave this implicit in much of the following for notational convenience. This also implies that if $L < (2+x^{-1})\Lambda$ then the whole integral is zero (new T-blocks created when the cutoff is $\Lambda$ have a minimum length $(2+x^{-1})\Lambda$). In Eq.~(\ref{4.2}) we denote this explicitly with a $\Theta$ function, but in this section we leave it implicit. The T-blocks that go into the TIT$\to$T moves are twice as numerous, and they are drawn according to their own distribution $p^T_\Lambda(\Lambda)$. The number density $n^T_\Lambda(\ell)$ therefore flows according to
\begin{eqnarray}\label{eqn:nT_flow}
\frac{\partial n^T_{\Lambda}(L)}{\partial \Lambda} =&& -2 n^I_\Lambda(\Lambda) p^T_\Lambda(L) \\ 
&& + n^I_\Lambda(\Lambda) \int_{\Lambda}^{\infty}d\ell p_{\Lambda}^{T}(\ell)p_{\Lambda}^{T}(L-\ell-x^{-1} \Lambda).\nonumber
\end{eqnarray}
%It is important to note that the last term in Eq.~(\ref{eqn:nT_flow}) is zero for $L<(2+x^{-1})\Lambda$ because $p^T_\Lambda(\ell)=0$ for $\ell<\Lambda$ (new T-blocks created when the cutoff is $\Lambda$ have a minimum length $[2+x^{-1}]\Lambda$). In Eq.~(\ref{4.2}) we denote this explicitly with a $\Theta$ function, but here we leave it implicit.

A similar line of reasoning for $n^I_\Lambda(d)$ and ITI$\to$I moves leads to
\begin{eqnarray}\label{eqn:nI_flow}
\frac{\partial n^I_{\Lambda}(D)}{\partial \Lambda} =&& -2 n^T_\Lambda(\Lambda) p^I_\Lambda(D) \\ 
&& +n^T_\Lambda(\Lambda) \int_{\Lambda}^{\infty}dd p_{\Lambda}^{I}(d)p_{\Lambda}^{I}(D-d+\Lambda).\nonumber
\end{eqnarray}

Noting that the number of blocks evolves according to $dN_\Lambda / d\Lambda = - [ n^T_\Lambda(\Lambda) + n^I_\Lambda(\Lambda) ]$ we can write Eqs.~(\ref{eqn:nT_flow}-\ref{eqn:nI_flow}) in terms of only $p^T_\Lambda(\ell)$ and $p^I_\Lambda(d)$, as was done in Refs.~\cite{Morningstar-Huse2019,Goremykina-Serbyn2018,Zhang-Huse2016}. As in Ref.~\cite{Morningstar-Huse2019}, the equation for $p^I_\Lambda(d)$ can be solved by an exponential form $\gamma_\Lambda \exp (- \gamma_\Lambda (d-\Lambda))$, where the parameter $\gamma_\Lambda$ flows with $\Lambda$. Another name for $\gamma_\Lambda$ here is $p^I_\Lambda(\Lambda)$, and we will use that going forwards. The flow equation that governs the evolution of this parameter $p^I_\Lambda(\Lambda)$ ends up being
\begin{eqnarray}\label{eqn:pILambda_flow}
\frac{dp^I_\Lambda(\Lambda)}{d\Lambda} = -p^T_\Lambda (\Lambda) p^I_\Lambda(\Lambda).
\end{eqnarray}
And the equation for $p^T_\Lambda(L)$ is
\begin{eqnarray}\label{eqn:pT_flow}
\frac{\partial p^T_{\Lambda}(L)}{\partial \Lambda} =&& [p^T_\Lambda(\Lambda)-p^I_\Lambda(\Lambda)]p^T_\Lambda(L) \\
&&+ p^I_\Lambda(\Lambda) \int_{\Lambda}^{\infty}d\ell p_{\Lambda}^{T}(\ell)p_{\Lambda}^{T}(L-\ell-x^{-1} \Lambda).\nonumber
\end{eqnarray}

Now we identify the rate $R_\Lambda = x p^I_\Lambda(\Lambda)$ because $x p^I(\Lambda) d\ell$ is the probability that an I-block has physical length $\in [\Lambda/x, \Lambda/x + d\ell ]$, and then $r_\Lambda(\ell) = x p^I_\Lambda (\Lambda) p^T_\Lambda(\ell)$ follows. Rewriting Eq.~(\ref{eqn:pT_flow}) now in terms of $R_\Lambda$ and $r_\Lambda(\ell)$ yields
\begin{eqnarray}
\frac{dr_\Lambda(L)}{d\Lambda} =&& \frac{1}{x}\left(\frac{dx}{d\Lambda} -R_\Lambda\right) r_\Lambda(L) \\ 
&& + \frac{1}{x}\int_\Lambda^{L-(1+x^{-1})\Lambda} d\ell r_\Lambda(\ell)r_\Lambda(L-\ell -x^{-1}\Lambda) \nonumber
\end{eqnarray}
as desired. This is an exact equation within the RG defined in Sec.~\ref{sec:concrete_RG}.

We now need to determine a flow equation for $x$. We do this by writing $x=(\sum_I d) / (\sum_I \ell)$, where $\sum_I$ denotes a sum over all I-blocks. Then again note that when the cutoff moves from $\Lambda$ to $\Lambda+d\Lambda$ the number of TIT$\to$T moves is $n_\Lambda^I(\Lambda)d\Lambda$ and the number of ITI$\to$I moves is $n_\Lambda^T(\Lambda)d\Lambda$. Therefore by the RG rules for combining blocks $(\sum_I d)_{\Lambda+d\Lambda} = (\sum_I d)_{\Lambda} - \Lambda n^T_\Lambda(\Lambda)d\Lambda - \Lambda n^I_\Lambda(\Lambda)d\Lambda$ and $(\sum_I \ell)_{\Lambda+d\Lambda} = (\sum_I \ell)_{\Lambda} + \Lambda n^T_\Lambda(\Lambda)d\Lambda - x^{-1} \Lambda n^I_\Lambda(\Lambda)d\Lambda$. The updated $x$ is then
\begin{equation}\label{eqn:x_update}
x_{\Lambda+d\Lambda} = \frac{(\sum_I d)_{\Lambda+d\Lambda}}{(\sum_I \ell)_{\Lambda+d\Lambda}}
= x_\Lambda - \frac{\Lambda n^T_\Lambda(\Lambda)(1+x)}{\sum_I \ell}d\Lambda.
\end{equation}
Now the distribution of $d$ (and $\ell = d/x$) is a known exponential, so $\sum_I \ell =N^I_\Lambda [1+\Lambda p_\Lambda^I(\Lambda)] / [xp^I_\Lambda(\Lambda)]$. Inserting this into Eq.~(\ref{eqn:x_update}), then using $n^T_\Lambda(\ell) = N^T_\Lambda p^T_\Lambda(\ell)$ and $N^T_\Lambda=N^I_\Lambda$ we arrive at
\begin{eqnarray}
\frac{dx}{d\Lambda} = - \frac{x(1+x)\Lambda p^I_\Lambda(\Lambda)p^T_\Lambda(\Lambda)}{1+\Lambda p^I_\Lambda(\Lambda)}.
\end{eqnarray}
Writing this in terms of rates results in
\begin{eqnarray}
\frac{dx}{d\Lambda} = - \frac{\Lambda r_\Lambda(\Lambda)(1+x)}{1+\Lambda R_\Lambda/x},
\end{eqnarray}
which is Eq.~(\ref{4.1}), as desired.

\section{Approximation of the convolution term\label{app:approximation}}

Here we analyze the integral (\ref{4.5}):
\begin{equation}\label{A.1}
\int d\ell^{\prime} d\ell
r_{\Lambda^{\prime}}(\ell)r_{\Lambda^{\prime}}(\ell^\prime),
\end{equation}
where $\Lambda^\prime = \Lambda + x(2\Lambda-\ell-\ell^\prime)$, and where the integration domain is restricted by the conditions $x\Lambda \le \Lambda^\prime \le \Lambda$, $\ell\ge \Lambda^\prime$, and $\ell^\prime \ge \Lambda^\prime$. We show that it is approximately equal to $R_\Lambda^2$.
The restrictions $x\Lambda \le \Lambda^\prime \le \Lambda$ translate to
\begin{equation} \label{A.2}
2\Lambda \le \ell +\ell^\prime \le \Lambda(1+x^{-1}).
\end{equation}
We also have the conditions $\ell, \ell^\prime \ge \Lambda^\prime = \Lambda +x(2\Lambda - \ell - \ell^\prime)$, which means that if either $\ell$ or $\ell^\prime$ is much smaller than $\Lambda$, then the other must be quite large. These boundaries of the integration region consist of rays going through $(\ell,\ell') = (\Lambda, \Lambda)$ with slopes $-x$ and $-1/x$. The integration domain is thus a triangle, as the lower limit in (\ref{A.2}) is superfluous. Observe that the entire square $\Lambda \le \ell, \ell^\prime \le \Lambda/(2x)$ is contained in the integration domain. 
Using Eq.~(\ref{independence}) to replace $r_{\Lambda^\prime}$ with $r_\Lambda$ in (\ref{A.1}), we see that
the integral over this square approximates well the full integral, $R_\Lambda^2$, as any contribution from $\ell$ or $\ell^\prime$ greater than $\Lambda/(2x)$ can be ignored. (Near the separatrix,  $r_{\Lambda^\prime}(\ell) \approx y_\Lambda/\ell^2$, which ensures that such contributions are down by a factor $x$, relative to the contribution from the square. Below the separatrix, the decay is even more rapid, as we will see in a moment.) We need to consider carefully the narrow wedges that expand the opening angle of the square to slightly more than $90^\circ$. To show their contributions are small, relative to $R_\Lambda^2$, consider first what happens near the separatrix, where $y$ is slowly varying and can be treated as a constant. 
Then using again Eq.~(\ref{independence}) and the fact that $\ell^\prime \ge \Lambda/2$ for $\ell \le \Lambda/(2x)$, we have that $r_{\Lambda^\prime}(\ell^\prime) = y_{\Lambda^\prime}/\ell^{\prime 2} \approx y_{\ell^\prime}/\ell^{\prime 2} \le 4y_{\ell^\prime}/\Lambda^2 \approx 4y_{\Lambda}/\Lambda^2 = 4r_\Lambda(\Lambda)$. Then the integral over one of the wedges is bounded by a constant times
\begin{eqnarray}
&& \int_\Lambda^{\Lambda/(2x)} d\ell r_\Lambda(\ell) r_\Lambda(\Lambda) x\ell \nonumber \\ 
&=& \int_\Lambda^{\Lambda/(2x)} d\ell \frac{y}{\ell^2}\frac{y}{\Lambda^2} x\ell \le x(\log x^{-1})\left(\frac{y}{\Lambda}\right)^2 \nonumber \\
&=& x(\log x^{-1})a^2R_\Lambda^2,
\end{eqnarray}
where $x\ell$ is a bound on the width of the wedge at $\ell$.
Near the separatrix, $a\approx 1$, so this case is complete. More generally, consider what happens anywhere below the separatrix. Again, from Eq.~(\ref{independence}), we have that $r_{\Lambda^\prime}(\ell) \approx r_\ell(\ell)$. 
Let us write $r(t) = r_\ell(\ell)$, where $\ell = e^t$, so that $r(t)=ye^{-2t}$. Then using Eq.~(\ref{a}), we have that $r^\prime/r = y^\prime/y - 2 = -(a+1)$, which leads to  an  exponential approximation 
\begin{align}\label{A.3}
r_\ell(\ell) =& r_\Lambda(\Lambda) \exp\left(\int_{\log \Lambda}^{\log \ell}(r^\prime/r) dt \right) \nonumber \\
=& r_\Lambda(\Lambda)\exp\left[-(a+1)\log (\ell/\Lambda)\right] \nonumber \\
\approx& r_\Lambda(\Lambda) \exp \left[ -(a+1)(\ell-\Lambda)/\Lambda \right],
\end{align}
for $(\ell-\Lambda)/\Lambda$ small.
Again using $\ell$ as the coordinate along the wedge, the density $r_{\ell^\prime}(\ell^\prime)$ acquires some growth in $\ell$ when $\ell^\prime$ is at the boundary, $\ell^\prime = \Lambda^\prime$. From Eq.~(\ref{A.3}), this leads to an extra factor of $\exp[ (a+1)x(\ell -\Lambda)/\Lambda ]$, due to the depression $\sim x(\ell-\Lambda)$ of $\Lambda^\prime$ below $\Lambda$. However, this is of little consequence, due to the decay of $r_\ell(\ell)$. In fact,  Eq.~(\ref{A.3}) effectively limits $\ell$ to the interval  $\ell-\Lambda \in [0,\Lambda/a]$, so the transverse integral can be estimated by $\text{density}\cdot \text{length} \approx (y_\Lambda/\Lambda^2)\cdot\frac{x\Lambda}{a} = xy_\Lambda/(\Lambda a) = xR_\Lambda$. The integral $d\ell$ along the wedge leads to another factor $R_\Lambda$, so this concludes the second case.
Note that we have not been allowing for the flow of $x$. However, since $dx/dt = -y$, it is clear this is also a small effect. In the end, one finds that the integral (\ref{A.1}) is very close to $R_\Lambda^2$, the errors down by a factor of at least $x\log x^{-1}$. 

\bibliography{main}

\end{document}